\documentclass[aps,prl,twocolumn,superscriptaddress,preprintnumbers,amsmath,amssymb]{revtex4}

\usepackage{graphicx} 
\usepackage{dcolumn}  
\usepackage{rotatin g}
\usepackage{booktabs}
\usepackage{SIunits} 
\usepackage{graphicx}
\usepackage{dcolumn}
\usepackage{diagbox}
\usepackage{bm}
\usepackage{color}
\usepackage[english]{babel}
\usepackage{titlesec}
\usepackage{multirow}
\usepackage[colorlinks,linkcolor=blue,urlcolor=blue,anchorcolor=green,citecolor=blue,breaklinks=true]{hyperref}
\usepackage{epstopdf}
\usepackage{enumitem}
\usepackage{hhline}
\usepackage{array}
\usepackage{threeparttable}
\usepackage{orcidlink}

\newcommand{\BR}{{\mathcal B}}

\begin{document} 
\hyphenpenalty=10000
\tolerance=1000
\vspace*{0.02\baselineskip}
  \author{I.~Adachi\,\orcidlink{0000-0003-2287-0173}} 
  \author{L.~Aggarwal\,\orcidlink{0000-0002-0909-7537}} 
  \author{H.~Ahmed\,\orcidlink{0000-0003-3976-7498}} 
  \author{H.~Aihara\,\orcidlink{0000-0002-1907-5964}} 
  \author{N.~Akopov\,\orcidlink{0000-0002-4425-2096}} 
  \author{A.~Aloisio\,\orcidlink{0000-0002-3883-6693}} 
  \author{N.~Althubiti\,\orcidlink{0000-0003-1513-0409}} 
  \author{N.~Anh~Ky\,\orcidlink{0000-0003-0471-197X}} 
  \author{D.~M.~Asner\,\orcidlink{0000-0002-1586-5790}} 
  \author{H.~Atmacan\,\orcidlink{0000-0003-2435-501X}} 
% ...
\collaboration{The Belle II Collaboration}

\preprint{\vbox{ 
	}
}

\title{ Search for the baryon number and lepton number violating decays $\tau^-\to \Lambda\pi^-$ and $\tau^-\to \bar{\Lambda}\pi^-$ at Belle II}

\noaffiliation

\begin{abstract}
We present a search for the baryon number $B$ and lepton number $L$ violating decays $\tau^- \rightarrow \Lambda \pi^-$ and $\tau^- \rightarrow \bar{\Lambda} \pi^-$ produced from the $e^+e^-\to \tau^+\tau^-$ process, using a 364 fb$^{-1}$ data sample collected by the Belle~II experiment at the SuperKEKB collider. 
No evidence of signal is found in either decay mode, which have $|\Delta(B-L)|$ equal to $2$ and $0$, respectively. Upper limits at 90\% credibility level on the branching fractions of $\tau^- \rightarrow \Lambda\pi^-$ and $\tau^- \rightarrow \bar{\Lambda}\pi^-$ are determined to be $4.7 \times 10^{-8}$ and $4.3 \times 10^{-8}$, respectively.
\end{abstract}

\maketitle
\tighten
\setcounter{footnote}{0}
Baryon number violation (BNV) is necessary to explain a dynamical generation of the asymmetry of matter and antimatter in the universe~\cite{Sakharov:1967dj, Weinberg:1981wj, Rubakov:1996vz, Morrissey:2012db}. For standard model (SM) processes, at perturbative level, lepton number $L$ and baryon number $B$ are conserved. However, nonperturbative effects, such as sphaleron processes, could lead to violations of $B$ and $L$ separately at high temperatures, while conserving their difference $B-L$~\cite{Phong:2020ybr, Papaefstathiou:2019djz, Zhou:2019uzq, Ho:2020ltr}. Several beyond-the-SM theories predict BNV, such as supersymmetry models with R-parity violation~\cite{Sjostrand:2002ip}, a black hole model~\cite{DeLuca:2021oer}, superstring models~\cite{Lazarides:1986th}, and grand unification models~\cite{deGouvea:2014lva}. Most of the models require $B-L$ conservation, while $|\Delta(B-L)|= 2$ is allowed in some scenarios~\cite{deGouvea:2014lva, Kamyshkov:1999yi, Wilczek:1979et}. Thus, BNV processes can be accompanied by lepton number violation. Discoveries of such BNV processes would reveal the existence of physics beyond the SM and shed light on matter-antimatter asymmetry.

Over the last few decades, several experiments have searched for BNV, but no evidence has been found~\cite{PDG}. Proton decays have been extensively studied by Super-Kamiokande~\cite{Super-Kamiokande:2009yit}, 
with lower limits on the lifetime of the proton on the order of $10^{33}$ years at the 90\% confidence level, while baryon-number-violating decays of charmed and bottom hadrons, $\tau$ leptons, and $Z$ bosons were studied by the CLEO~\cite{CLEO:2009apb, CLEO:1999emi}, BaBar~\cite{BaBar:2011yks, BaBar:2011ouc}, OPAL~\cite{OPAL:1998gbn}, Belle~\cite{Belle:2005exq, Belle:2023mao}, and LHCb Collaborations~\cite{LHCb:2013fsr}, with upper limits on the branching fractions in the range $10^{-8} - 10^{-5}$ at the 90\% confidence level.

We use a sample of $e^+e^-$ collisions collected at a center-of-mass (c.m.)\ energy $\sqrt{s}$ of 10.58 GeV corresponding to the mass of the $\Upsilon(4S)$ resonance to search for BNV decays $\tau^- \rightarrow \Lambda \pi^-$ and $\tau^- \to \bar{\Lambda} \pi^-$. The data, recorded by the Belle II detector~\cite{Belle-II:2010dht} operating at the SuperKEKB~\cite{AKAI2018188} asymmetric-energy collider, have an integrated luminosity $\mathcal{L}$ of ($364 \pm 2$) fb$^{-1}$~\cite{Lum}. Searches for $\tau^- \rightarrow \Lambda \pi^-$ and $\bar{\Lambda} \pi^-$ were performed by Belle~\cite{Belle:2005exq} using a 154 fb$^{-1}$ data sample yielding upper limits of $0.72 \times 10^{-7}$ and $1.4 \times 10^{-7}$, respectively, at the 90\% confidence level. With an upgraded detector and a larger data sample, the Belle II experiment can investigate these decays with improved sensitivity.

We select $e^+e^-\rightarrow \tau^+\tau^-$ events in which one of the two $\tau$ leptons decays into three charged particles (3-prong) and the other into one charged particle (1-prong) and one or more neutrinos. We search for a signal on the 3-prong (signal) side via the decay $\tau^- \to \Lambda(\to p \pi^-)\pi^- $ or $\tau^- \to \bar{\Lambda}(\to \bar{p} \pi^+)\pi^-$, while we require that the other $\tau$ (tag) decays into $e^+ \nu_e \bar{\nu}_{\tau}$, $\mu^+ \nu_{\mu} \bar{\nu}_{\tau}$, $\pi^+ \bar{\nu}_{\tau}$, or $\pi^+ \pi^0 \bar{\nu}_{\tau}$ final states.
Inclusion of charge-conjugate states is always implied. A search region is defined in a plane formed by the energy difference $\Delta E = E(\Lambda\pi) - \sqrt{s}/2$ and the invariant mass $M(\Lambda\pi)$, where $\Lambda$ means $\Lambda$ or $\bar{\Lambda}$, 
and signal is identified as an excess over background expectations.

In the following, all variables are defined in the $e^+e^-$ c.m.\ system unless otherwise specified. The analysis is optimized using simulated events, and data events outside the signal region, before examining data in the signal region.


The Belle II detector has a cylindrical geometry with the axis of symmetry along the beamline ($z$ axis)~\cite{Belle-II:2010dht}. The innermost component is a two-layer silicon-pixel detector surrounded by a four-layer double-sided silicon-strip detector and a 56-layer central drift chamber (CDC). These detectors reconstruct trajectories of charged particle (tracks). Only one sixth of the second layer of the silicon-pixel detector was installed for the data analyzed here. Surrounding the CDC, which also provides d$E$/d$x$ energy-loss measurements, is a time-of-propagation detector in the central region and an aerogel-based ring-imaging Cherenkov detector in the forward region. These detectors provide particle identification (PID) for charged hadrons. Surrounding them is an electromagnetic calorimeter (ECL) composed of CsI(Tl) crystals that primarily provides energy and timing measurements for photons and electrons. Outside of the ECL is a superconducting solenoid magnet. Its flux return is instrumented with resistive-plate chambers and plastic scintillator modules to detect muons, $K^0_L$ mesons, and neutrons. The solenoid magnet provides a 1.5~T magnetic field parallel to the $z$ axis. The longitudinal direction and the polar angle $\theta$ are defined with respect to the $z$ axis, whose positive direction is that of the electron beam.


Simulated samples are used to estimate signal efficiencies and the number of expected background events~\cite{Zhou:2020ksj}.
We use the {\sc KKMC} software package to generate $2 \times 10^7$ $e^+e^- \rightarrow \tau^+\tau^-(\gamma)$ signal events~\cite{JADACH2000260}.
The subsequent $\tau$ decays are simulated by the {\sc TAUOLA} software package~\cite{JADACH1991275}, with one $\tau$ decaying to $\Lambda\pi$ or $\bar{\Lambda}\pi$ according to a phase-space model and the other decaying according to known branching fractions.~\cite{PDG, JADACH2000260}. We use {\sc KKMC} and {\sc TAUOLA} also to simulate backgrounds.
The {\sc KKMC} generator is also used for $e^+e^- \rightarrow \mu^+\mu^-(\gamma)$ and $q\bar{q}$ processes, where $q$ indicates $u$, $d$, $s$, or $c$ quarks, with the $q\bar{q}$ pair fragmentation being simulated by the {\sc PYTHIA8} software package~\cite{SJOSTRAND2015159}.
The {\sc PYTHIA8} and {\sc EvtGen}~\cite{Lange:2001uf} software packages are used to simulate the $e^+e^- \to b\bar{b}$ process. The {\sc BabaYaga@NLO} software package is used to simulate the $e^+e^- \rightarrow e^+e^-(\gamma)$ processes~\cite{Balossini:2006wc, Balossini:2008xr, CarloniCalame:2003yt, CarloniCalame:2001ny, CarloniCalame:2000pz}; the {\sc AAFH}~\cite{Berends:1984ge, Berends:1984gf, Berends:1986ig} and {\sc TREPS}~\cite{Uehara:1996bgt} software packages are used for simulations of $e^+e^- \rightarrow \ell^+\ell^-\ell^+\ell^-$ and $e^+e^-h^+h^-$ processes, respectively, where $\ell$ indicates an electron, muon, or $\tau$ lepton, and $h$ indicates a pion, kaon, or proton.
The size of each simulated sample is four times the size of the corresponding component in data, except for the $e^+e^- \to \ell^+\ell^-\ell^+\ell^-$ and $e^+e^-h^+h^-$ samples, which are the same size as expected in data, and the $e^+e^- \to e^+e^-(\gamma)$ sample, which is 10\% of that expected in data.
{\sc EvtGen} is used to simulate the decay of hadrons~\cite{Lange:2001uf}. 
The {\sc PHOTOS} software package is used for the simulation of final state radiation~\cite{Barberio:1990ms}. 
The Belle II software~\cite{Kuhr:2018lps} uses the {\sc Geant4}~\cite{GEANT4:2002zbu} software package to simulate the response of the detector to the interactions of particles. 

The hardware trigger relies on energy deposits (clusters with energy larger than 100 MeV) and their topologies in the ECL, and on the number of reconstructed tracks in the CDC. Events are required to satisfy one of the following criteria: three clusters with a topology inconsistent with a Bhabha process, where at least one cluster has an energy greater than 500 MeV;
the sum of clusters exceeding 1 GeV; at least one charged particle with a momentum greater than 0.7 GeV/$c$.
The trigger efficiency for the $\tau^- \to \Lambda \pi^-$ $(\tau^- \to \bar{\Lambda} \pi^-)$ channel, estimated on simulated signal samples, is $99.5\%$ $(99.6\%)$.


Both charged and neutral particles are required to be within the acceptance of the CDC, i.e., \mbox{$-0.866 < \cos\theta < 0.956$}. The transverse momentum of each charged particle is required to exceed \mbox{0.1 GeV/$c$}. Photons are identified as clusters with energies greater than 0.1 GeV
not associated to any track. The thrust axis $\hat{t}$ is defined such that the value
\begin{equation}
    \label{eq1}
	T=\frac{\sum_i \left|\vec{p} \, _i \cdot \hat{t} \right|}{\sum_i \left|\vec{p} \, _i\right|}
\end{equation}
is maximized. Here, $\vec{p}\,_i$ is the momentum of the $i$th particle and the sum runs over all tracks and clusters. Events are geometrically split into two opposite hemispheres with a plane perpendicular to the thrust axis. The hemispheres corresponding to the signal and tag sides are required to contain exactly three tracks and one track, respectively. In the following, we refer to the pion from the $\Lambda$ or $\bar{\Lambda}$ decays as $\pi_{\Lambda}$, the pion from the signal $\tau$ decays as $\pi_{\tau}$, and the pion from the tag $\tau$ decays as $\pi_{\rm tag}$. 
We require that the numbers of photons $N_{\gamma}^{\text{sig}}$ and $N_{\gamma}^{\text{tag}}$ on the signal and tag sides each be less than four, and that
the energy $E_{\gamma}^{\text{sig}}$ of the most energetic photon on the signal side be less than 1.0 GeV. These requirements take into account the possibility of photons radiated from the initial state and photons from $\pi^0$ decays on the tag side.

\begin{figure}[htbp]
    \centering
    \includegraphics[width=8.7cm]{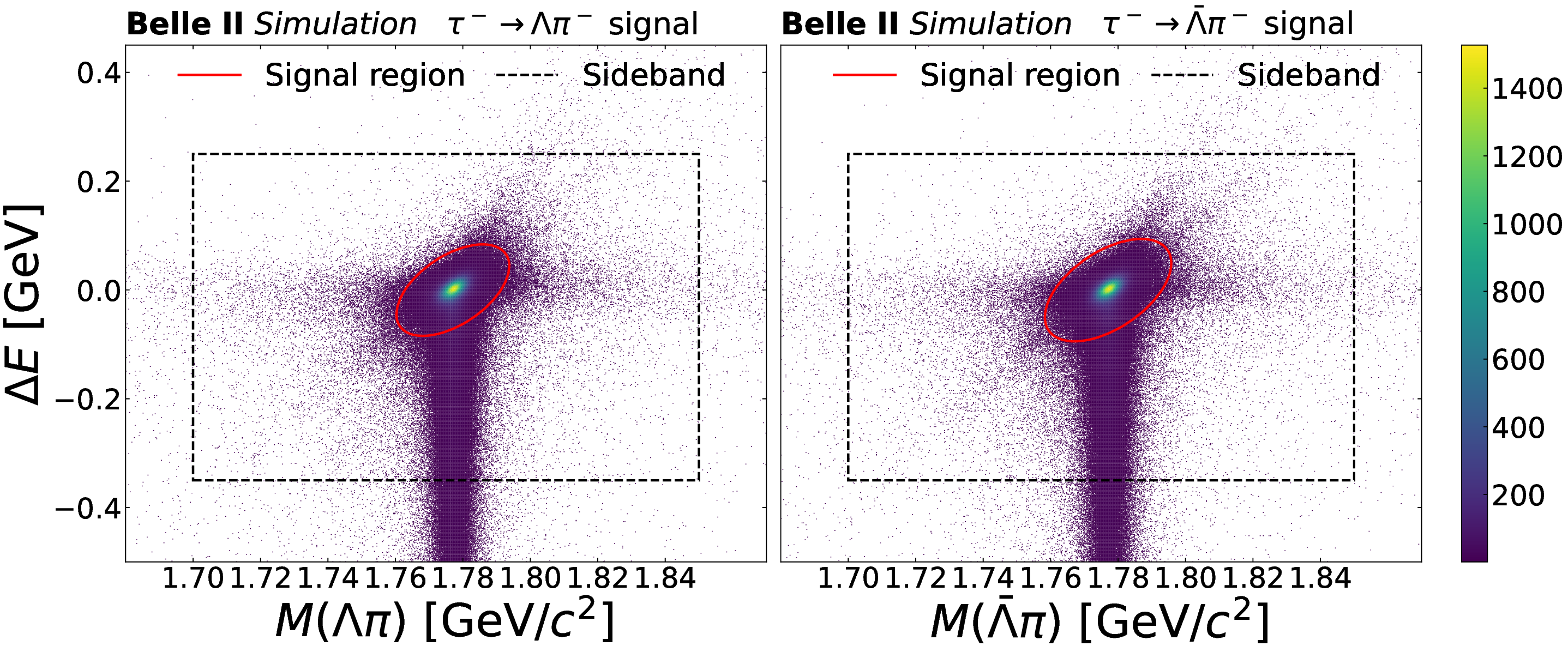}
    \put(-147,81){\bf (a)}
    \put(-45,81){\bf (b)}
    \caption{Distributions of $\Delta E$ as a function of $M(\Lambda\pi)$ for simulated (a) $\tau^- \rightarrow \Lambda \pi^-$ and (b) $\tau^- \rightarrow \bar{\Lambda} \pi^-$ signal samples. The red solid ellipses identify the signal regions, while the areas between the dashed black boxes and the corresponding ellipses are the sideband regions.}
    \label{MDE2DSig}
    
\end{figure}

A signal region is defined in the plane of the $\Delta E$ versus $M(\Lambda\pi)$, in which signal is expected to peak at $\Delta E \approx$ 0.0 and at $M(\Lambda\pi) \approx$ 1.78 GeV/$c^2$.
The signal region, shown in Fig.~\ref{MDE2DSig}, has an elliptical shape whose size is defined so as to include 90\% of all reconstructed signal events. The distribution of signal in the $\Delta E$ versus $M(\Lambda\pi)$ plane is broadened by detector resolution and radiative effects. Photons radiated from the intial state lead to a tail at low values of $\Delta E$. A sideband region is also defined as the complements of the signal regions enclosed in the ranges \mbox{$1.70$ GeV/$c^2 < M(\Lambda\pi) < 1.85$ GeV/$c^2$} and \mbox{$-0.35$ GeV $< \Delta E < 0.25$ GeV}, indicated by the rectangular boxes in Fig.~\ref{MDE2DSig}. Only candidates falling within the rectangular boxes in Fig.~\ref{MDE2DSig} are analyzed.

We require that charged particles on the signal and tag sides be identified by combining information from various subdetectors to form the likelihood ${\cal L}_i$ for particle species $i$. The ratio \mbox{${\cal{R}}_{i}$ = ${\cal{L}}_{i}/{\Sigma_{j}{\cal{L}}_{j}}$} is used to identify protons and muons, where $i$ indicates proton or muon, and $j$ indicates electron, moun, pion, kaon, proton, and deuteron. We require ${\cal{R}}_{p}>$ 0.5 and ${\cal{R}}_{\mu}>$ 0.9, with efficiencies of 98.5\% and 95\%, respectively; 3.4\% (6.5\%) probability for misidentifying a proton as a pion (kaon); and 4.1\% for misidentifying a muon as a pion.
The electron identification relies on a boosted decision tree classifier trained with information from all subdetectors except the silicon vertex detectors.
We require the boosted decision tree output ${\cal{P}}^{e}>$ 0.9, with an efficiency of 99.4\%, and the rate for misidentifying an electron as a pion is 0.7\%. 
Pion identification relies on the ratio for particle types $m$ and $m^{\prime}$, ${\cal{R}}(m|m^{\prime})$ = ${\cal{L}}_{m}/({{\cal{L}}_{m}+{\cal{L}}_{m^{\prime}}})$, where $m$ and $m^{\prime}$ represent proton, pion, and kaon. For $\pi_{\tau}$ and $\pi_{\rm tag}$, we require ${\cal{R}}(p|\pi)<$ 0.6 and ${\cal{R}}(K|\pi) <$ 0.4, with identification efficiencies of 98.2\% and 99.9\%, and the rates for misidentifying a pion as a proton or a pion as a kaon are 9.4\% and 8.6\%, respectively. The charged particle on the tag side is required to be identified as $e$, $\mu$, or $\pi$.

\begin{figure}[htbp]
    \centering
    \includegraphics[width=8.9cm]{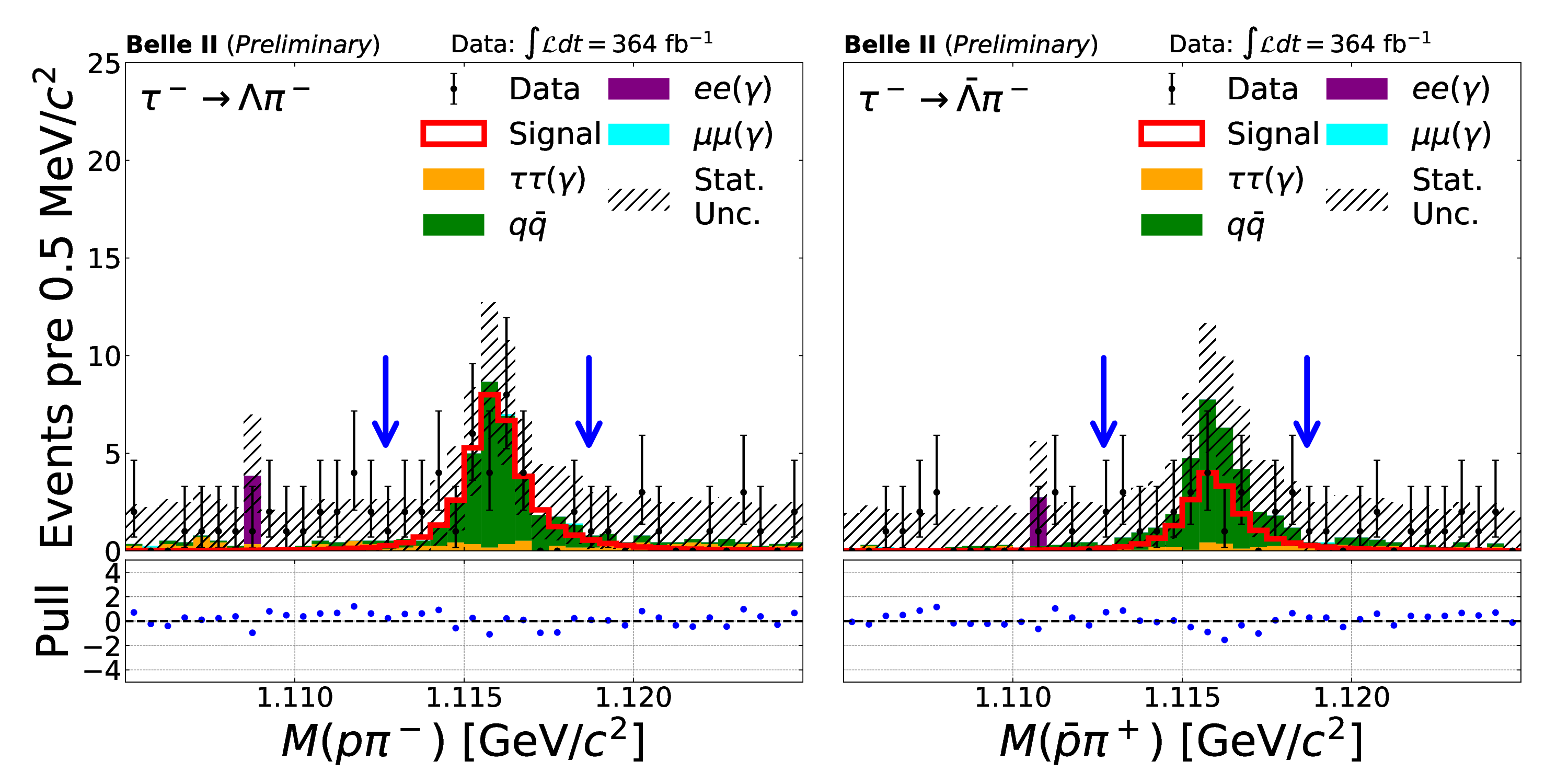}
    \put(-230,97){\bf (a)}
    \put(-115,97){\bf (b)}
    \\
    \includegraphics[width=8.9cm]{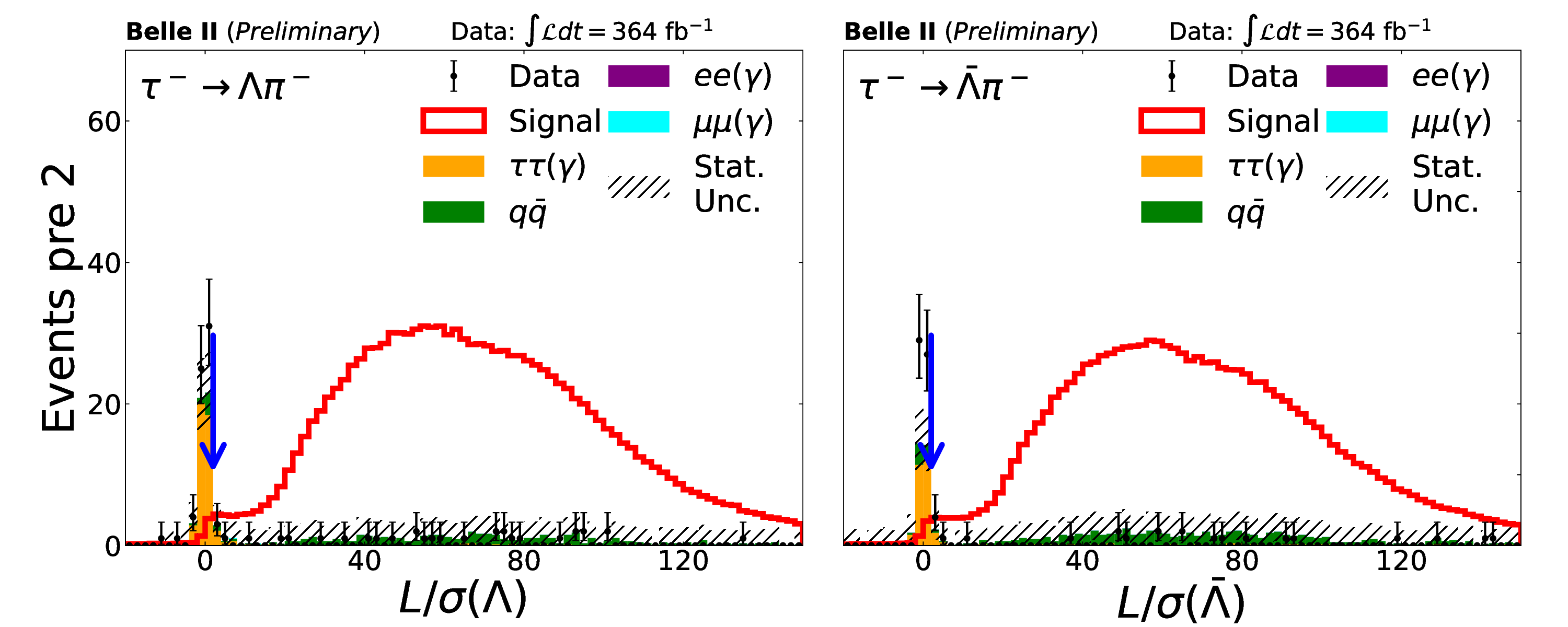}
    \put(-230,75){\bf (c)}
    \put(-115,75){\bf (d)}
    \caption{Distributions of (a) $M(p\pi^-)$ for $\tau^- \rightarrow \Lambda \pi^-$ candidates, (b) $M(\bar{p}\pi^+)$ for $\tau^- \rightarrow \bar{\Lambda} \pi^-$ candidates, (c) $L/\sigma$ of $\Lambda$ candidates for $\tau^- \rightarrow \Lambda \pi^-$ and (d) $L/\sigma$ of $\bar{\Lambda}$ candidates for $\tau^- \rightarrow \bar{\Lambda} \pi^-$. 
    The red open histograms show the simulated signal distributions, the filled histograms are stacked to show the simulated background distributions, with statistical uncertainties displayed as hatched areas, and the points with error bars show the distributions of the data in the sideband regions. The simulated signal distribution is arbitrarily scaled. The blue arrows indicate the boundaries of the selection criteria. Pull distributions show the difference between data and simulation divided by the expected uncertainty on the model.}
    \label{Lambda}
\end{figure}

We reconstruct $\Lambda$ and $\bar{\Lambda}$ candidates by combining an identified $p$ or $\bar{p}$ with an oppositely charged particle for which no particle identification is required. The resulting $p \pi^-$ and $\bar{p}\pi^+$ invariant mass distributions are shown in Figs.~\ref{Lambda}(a) and~\ref{Lambda}(b). Candidates within a $6$ MeV/$c^2$ range around the known $\Lambda$ mass are selected~\cite{PDG}. The flight significance of $\Lambda$ and $\bar{\Lambda}$ candidates, defined as the ratio $L/\sigma$ of the flight distance $L$ to its uncertainty $\sigma$, is required to be larger than 2.0, as shown in Figs.~\ref{Lambda}(c) and~\ref{Lambda}(d). 
After applying these selections, 83\% (88\%) of the remaining candidates are correctly identified as $\Lambda$ ($\bar{\Lambda}$), while the remaining 17\% (12\%) are due to random combinations of tracks, according to simulation.

The missing momentum $p_{\rm miss}$ is defined as the difference between the total initial momentum and the sum of the momenta of all charged particles and photons. To suppress backgrounds from $e^+e^- \to e^+e^-(\gamma)$, $e^+e^- \to \mu^+\mu^-(\gamma)$, $e^+e^- \to \ell^+\ell^-\ell^+\ell^-$, and $e^+e^- \to e^+e^-h^+h^-$ processes, the cosine of the angle between $p_{\rm miss}$ and the track on the tag side is required to be greater than 0.15. The angle $\theta_{\Lambda-\pi_{\tau}}$ between the momenta of the $\Lambda$ candidates and $\pi_{\tau}$ is required to be greater than 0.1 radians. 
We also require $T > 0.9$. After these selections are applied, contributions from $e^+e^- \to \ell^+\ell^-\ell^+\ell^-$ and $e^+e^- \to e^+e^-h^+h^-$ processes are negligible. At this stage of the analysis, 99.2\% (99.3\%) of the background is removed, with a signal efficiency of 74.3\% (74.7\%) for the $\tau^- \to \Lambda (\bar{\Lambda}) \pi^-$ decay modes. The remaining background is dominated by $e^+e^-\to \tau^+\tau^-(\gamma)$ and $e^+e^- \to q\bar{q}$ processes.

To further suppress the remaining background, a gradient boosted decision tree (GBDT) classifier 
is used~\cite{PSpeckmayer_2010}. We use 15 discriminating observables defined at the event level, on the signal side, and on the
tag side as inputs
to the classifier.
The observables at event level are the sum of the energies of all visible particles $E_{\rm vis}$; the magnitude of the missing momentum; the square of the missing mass $M^{2}_{\rm miss} = E^{2}_{\rm miss} - p^{2}_{\rm miss}$, where $E_{\rm miss}$ is the difference between the total initial energy and $E_{\rm vis}$; the cosine of the polar angle of the missing momentum; the cosine of the angle between the missing momentum and $\hat{t}$; and the cosine of the angle between the missing momentum and the $\Lambda\pi$ ($\bar{\Lambda}\pi$) system. The observables related to the signal side are $N^{\rm sig}_{\gamma}$; $E_{\gamma}^{\rm sig}$; $\theta_{\Lambda-\pi_{\tau}}$; the angle between the momenta of $\pi_{\tau}$ and $\pi_{\Lambda}$; and the momentum of the $\Lambda\pi$ ($\bar{\Lambda}\pi$) system. The observables related to the tag side are $N^{\rm tag}_{\gamma}$; the energy of the most energetic photon on the tag side; the mass of the system recoiling against the track on the tag side; and an identification code assigned to the track’s particle-identification information to distinguish between $e$, $\mu$, and $\pi$. We train the GBDT classifier with samples of simulated signal and background events satisfying the selections described above. The GBDT for the two decay channels are trained separately. Signal and background events are divided into training and test samples in a 1:4 ratio. 

\begin{figure}[htbp]
    \centering
    \includegraphics[width=8.9cm]{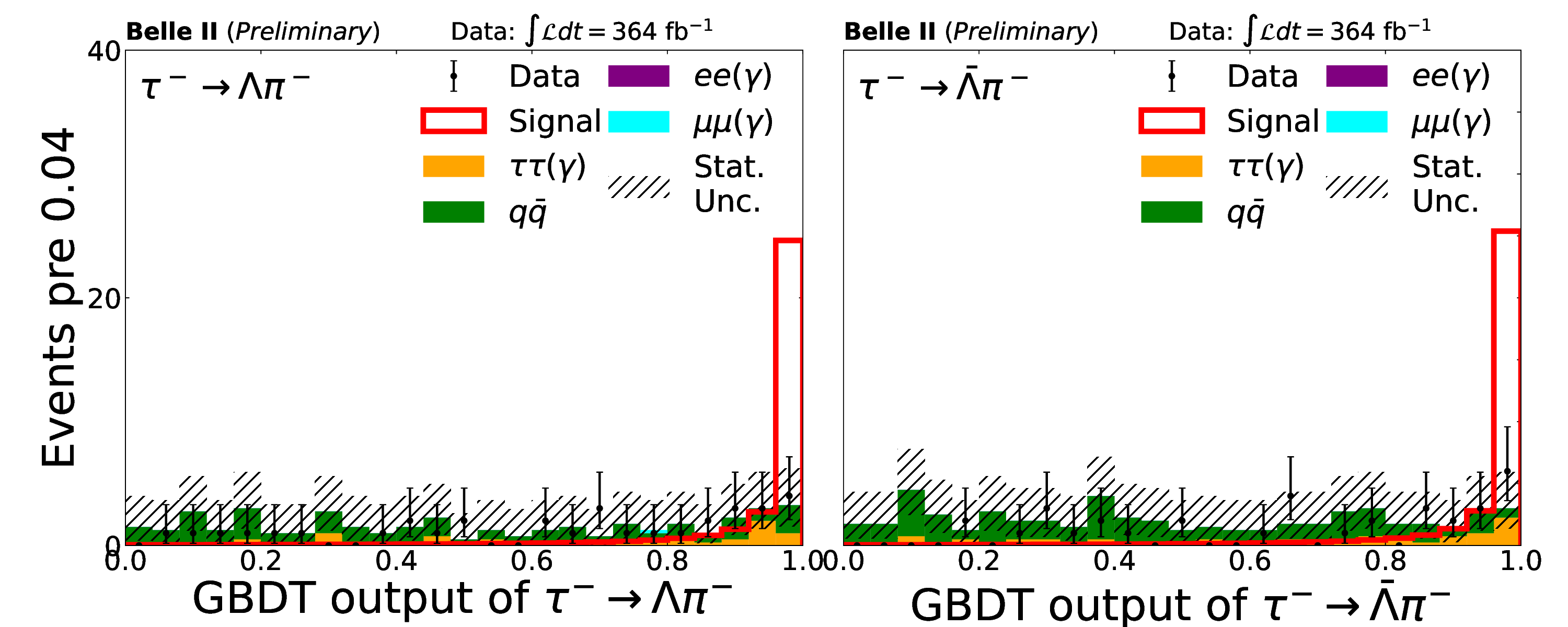}
    \put(-230,75){\bf (a)}
    \put(-112,75){\bf (b)}
    \caption{Distributions of GBDT outputs for (a) $\tau^- \rightarrow \Lambda\pi^-$ and (b) $\tau^- \rightarrow \bar{\Lambda}\pi^-$ samples. The red open histograms show the simulated signal distributions, the filled histograms are stacked to show the simulated background distributions, with statistical uncertainties displayed as hatched areas, and the points with error bars show the distributions of the data in the sideband regions. The simulated signal distribution is arbitrarily scaled.}
    \label{BDTG}
\end{figure}

The distributions of the GBDT outputs are shown in Fig.~\ref{BDTG}. To optimize the requirements on the GBDT output values, we minimize the expected upper limits at 90\% credibility level~\cite{CL1} on $\BR(\tau^- \rightarrow \Lambda \pi^-)$ and $\BR(\tau^- \rightarrow \bar{\Lambda} \pi^-)$ (as described later in the paper) estimated with a simulated sample independent of those used for training and testing.
The optimized selections are ${\cal{P}}^{\text{GBDT}} > 0.886$ $(0.803)$, resulting in a 78.8\% (82.3\%) relative signal efficiency due to the GBDT only and 98.9\% (98.8\%) background rejection for the $\tau^- \rightarrow \Lambda \pi^-$ $(\tau^- \rightarrow \bar{\Lambda} \pi^-)$ decay mode.  These values are consistent for both the training and test samples. The final signal efficiencies are 9.5\% and 9.9\% for $\tau^- \rightarrow \Lambda\pi^-$ and $\tau^- \rightarrow \bar{\Lambda} \pi^-$ decays, respectively, which include corrections for PID, trigger, and $\Lambda$ selection, as described below.

The branching fraction of the $\tau^- \rightarrow \Lambda \pi^-$ decay channel is calculated using 
\begin{equation}
  \BR(\tau^- \rightarrow \Lambda \pi^-) = \frac{N_{\rm sig}}{2 \varepsilon_{\rm sig} \mathcal{L} \sigma_{\tau\tau} \BR(\Lambda \rightarrow p \pi^-)},
  \label{eq4}
\end{equation}
where $N_{\rm sig}$ is the signal yield, $\varepsilon_{\rm sig}$ is the signal efficiency, $\mathcal{L}$ is the integrated luminosity, $\sigma_{\tau\tau} = 0.919 \pm 0.003$ nb is the cross section of $e^+e^- \rightarrow \tau^+\tau^-(\gamma)$~\cite{Banerjee:2007is}, and $\BR(\Lambda \rightarrow p \pi^-)=0.641\pm0.005$ is the known branching fraction of $\Lambda \rightarrow p \pi^-$~\cite{PDG}. This formula also applies to the $\tau^- \rightarrow \bar{\Lambda} \pi^-$ decay mode.

Several corrections and systematic uncertainties affect the terms in Eq.~\ref{eq4} and therefore the determination of the branching fraction. The sources are from signal efficiency, luminosity, $\sigma_{\tau\tau}$, $\BR(\Lambda \rightarrow p \pi^-)$, and signal yield.
\par
Sources of systematic uncertainties for the signal efficiency include uncertainties in the efficiencies of the $\Lambda$ selection, GBDT selection, tracking, PID, and trigger.
The uncertainty due to the requirement on the flight significance for the $\Lambda$ selection is studied using a $\Lambda_c^{+} \to \Lambda \pi^+$ control channel. The ratio of $\Lambda$ selection efficiencies in data and simulation is $0.977 \pm 0.002$ \mbox{$(0.976 \pm 0.002)$} for $\tau^- \to \Lambda(\bar{\Lambda}) \pi^-$ channel. These values are used to correct the signal efficiencies, and the corresponding uncertainties are taken as systematic uncertainties.
The systematic uncertainties on the signal efficiency from the GBDT selection are studied by splitting the training samples into two equal size parts and training the GBDTs separately. The differences between the resulting signal efficiencies after applying the same requirements on the output values as in the nominal analysis are both 0.50\% for the two decay channels, which are taken as systematic uncertainties due to the GBDT.
The uncertainty due to tracking efficiency is 0.24\% per track, estimated with $e^+e^- \rightarrow \tau^+\tau^-$ events, leading to a systematic uncertainty of 0.96\%. 
The systematic uncertainties from charged-particle identification are studied in $\Lambda \rightarrow p \pi$, $D^{*+} \rightarrow D^0(\rightarrow K^-\pi^+)\pi^+$, and $K_S^0 \rightarrow \pi^+ \pi^-$ decays for protons and pions, and in $J/\psi \rightarrow e^+e^-(\gamma)$, $\mu^+\mu^-(\gamma)$, $e^+e^- e^+e^-$, and $e^+e^- \mu^+\mu^-$ processes for muons and electrons.
Correction factor to the signal efficiency for the $\tau^- \to \Lambda (\bar{\Lambda})\pi^-$ decay mode is $0.953\pm0.021$ ($0.954\pm0.022$) for the $\tau^- \to \Lambda (\bar{\Lambda})\pi^-$ channel.
The quadratic sum of the individual charged-particle contributions is taken at the systematic uncertainty from charged particle identification.
The systematic uncertainty on the trigger efficiency is studied using $e^+e^- \to \tau^+ \tau^-$ events with $\tau^- \to \pi^-\pi^+\pi^- \nu_{\tau}$ and $\tau^+$ decaying into the same final states as in the tag side of this analysis~\cite{Tau3mu}. The trigger efficiency measured in data differs by 0.7\% from that in simulation; this difference is taken as the systematic uncertainty due to the trigger. 
\par
The uncertainty in the branching fraction $\mathcal{B}(\Lambda \rightarrow p \pi^-)$ is 0.78\%~\cite{PDG}. 
The uncertainty of the $e^+e^- \to \tau^+\tau^-(\gamma)$ cross section, estimated with {\sc KKMC}, is 0.33\%~\cite{Banerjee:2007is}. The uncertainty in the integrated luminosity is 0.60\%, determined from the large-angle Bhabha scattering process~\cite{Belle-II:2019usr}. The quadratic sum of all the above uncertainties, summarized in Table~\ref{SysTotal}, is \mbox{$\sigma$ = 2.77\% (2.82\%)} for the $\tau^- \rightarrow \Lambda(\bar{\Lambda})\pi^-$ channel.
\begin{table}[htbp]
    \caption{\label{SysTotal} Summary of fractional systematic uncertainties.}
    \renewcommand\arraystretch{1.2}
    \begin{tabular}{l c c}
    \hline\hline
    ~~\multirow{2}*{Source}~~ & \multicolumn{2}{c}{Uncertainty (\%)} \\
    \cline{2-3}
    ~~ &$\tau^-\to \Lambda \pi^-$     & $\tau^-\to \bar{\Lambda} \pi^-$ \\\hline
	$\Lambda$ selection						   & 0.20   & 0.20\\
        GBDT selection	               			   & 0.50   & 0.50\\
	Tracking efficiency                        & 0.96   & 0.96\\
        Particle identification                    & 2.21   & 2.28\\
        Trigger efficiency                         & 0.70   & 0.70\\
	$\mathcal{B}(\Lambda \rightarrow p \pi^-)$ & 0.78   & 0.78\\
	$\tau$-pair cross section                  & 0.33   & 0.33\\
	Luminosity                                 & 0.60   & 0.60\\
        \hline
        Total                                      & 2.77   & 2.82\\
    \hline\hline
    \end{tabular}
\end{table}

\par
The estimate of $N_{\rm sig}$ (see Eq.~\ref{eq4}) depends on the expected background yield. Systematic uncertainties affecting the expected background yield and due to data-simulation discrepancies are evaluated from the sideband regions.
There are $N_{\rm SB}^{\rm data} = 7$ $(6)$ data events in the sideband region for the $\tau^- \rightarrow \Lambda(\bar{\Lambda}) \pi^-$ channel, while the corresponding number of events observed in simulated background samples is \mbox{$N_{\rm SB}^{\rm sim} = 3.2^{+1.7}_{-1.2}$ $(5.5^{+2.1}_{-1.6})$}, where the uncertainties are due to the limited size of the simulated samples. The data-simulation ratios $f_{\rm bkg} = N_{\rm SB}^{\rm data}/N_{\rm SB}^{\rm sim}$ are used as correction factors for the background estimated in the signal region: they are $2.2^{+1.7}_{-1.2}$ ($1.1^{+0.8}_{-0.5}$) for the $\tau^- \rightarrow \Lambda(\bar{\Lambda})\pi^-$ channels, where the uncertainties are statistical only. 
The relative uncertainties in $f_{\rm bkg}$ are treated as systematic uncertainties on the expected background yield, being $^{+77\%}_{-55\%}$ ($^{+73\%}_{-45\%}$) for the $\tau^- \rightarrow \Lambda(\bar{\Lambda}) \pi^-$ channel. Some of the sources of uncertainty listed in Table~\ref{SysTotal} also affect the expected background yield. Their contributions are negligible compared to those affecting the correction factors estimated in the sidebands and are not taken into account.

\begin{figure}[htbp]
    \centering
    \includegraphics[width=8cm]{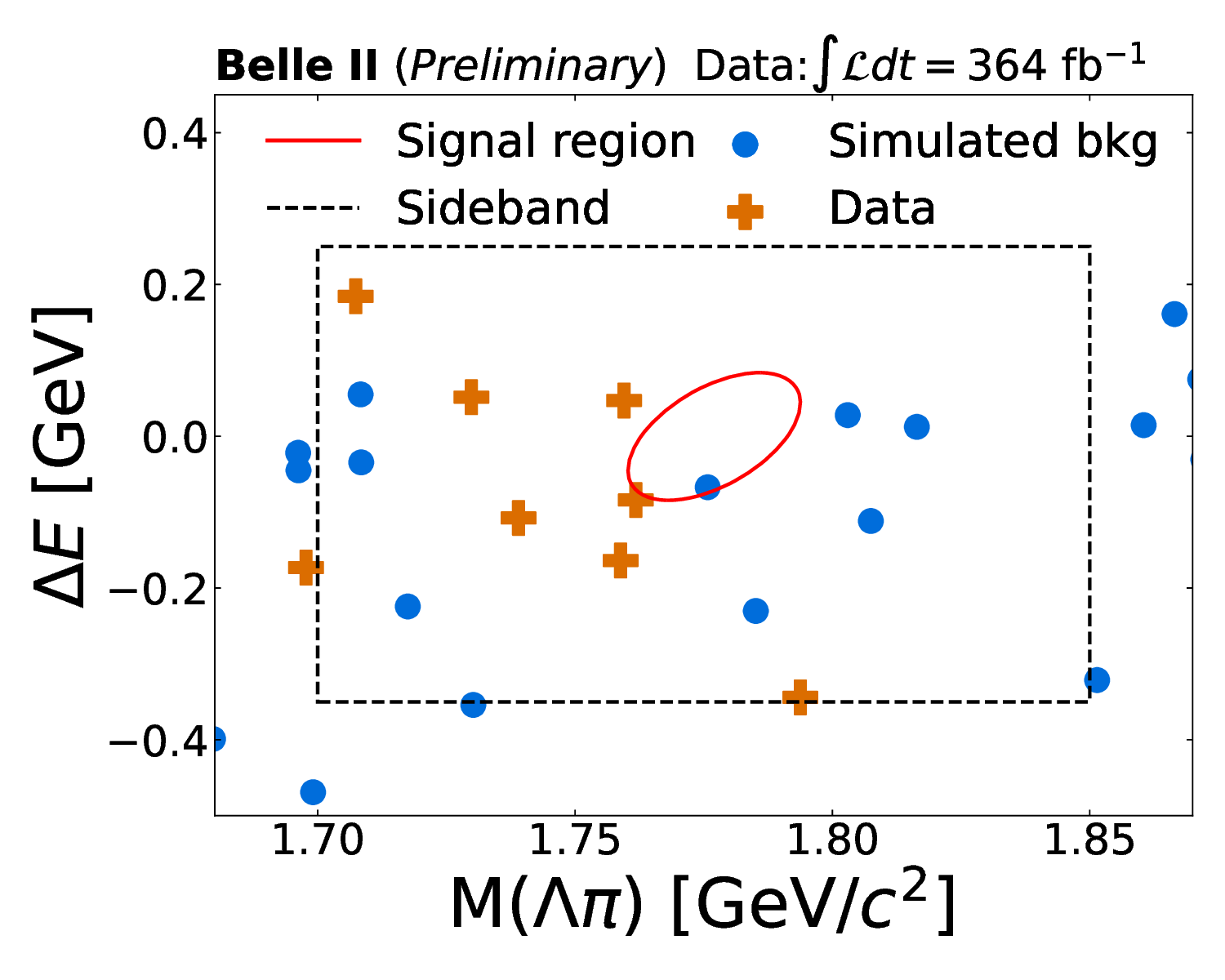}
    \put(-25,140){\bf (a)}
    \\
    \includegraphics[width=8cm]{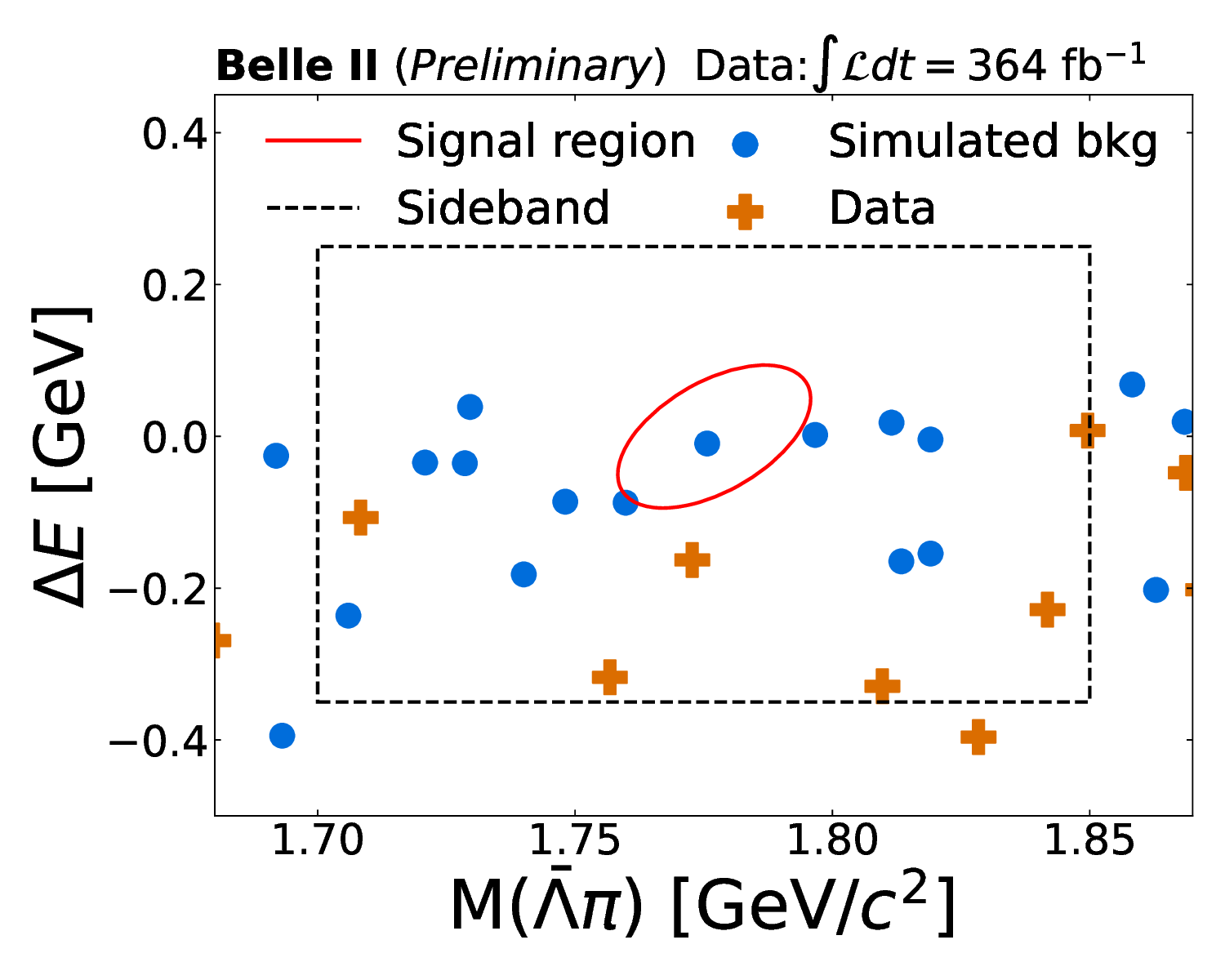}
    \put(-25,140){\bf (b)}
    \caption{Distributions of the events for data and simulated background in the $\Delta E$ versus $M(\Lambda\pi)$ plane for (a) $\tau^- \rightarrow \Lambda \pi^-$ and (b) $\tau^- \rightarrow \bar{\Lambda} \pi^-$ candidates after applying all selections.}
    \label{MDE2D}
\end{figure}

The distributions of the events in the $\Delta E$ versus $M(\Lambda\pi)$ plane are shown in Fig.~\ref{MDE2D}. From simulation, the expected background yields for $\tau^- \rightarrow \Lambda\pi^-$ and $\tau^- \rightarrow \bar{\Lambda}\pi^-$ within the signal region, after applying the correction factor $f_{\rm bkg}$, are $N_{\rm exp} = 1.0^{+1.3}_{-1.1}$ and $0.5\pm0.6$, respectively, where the uncertainties include both statistical and systematic uncertainties. No events are observed in the signal region. The branching fractions are measured to be $(-2.5^{+4.1+1.9}_{-3.7-1.4})\times10^{-8}$ and $(-1.2\pm2.8^{+0.9}_{-0.5})\times10^{-8}$ for the $\tau^- \rightarrow \Lambda\pi^-$ and $\tau^- \rightarrow \bar{\Lambda}\pi^-$ channels, respectively, where the first uncertainties are statistical and the second are systematic.

Since no signal is observed, we compute upper limits on the signal yields for the $\tau^- \rightarrow \Lambda \pi^-$ and $\tau^- \rightarrow \bar{\Lambda} \pi^-$ channels. We assume that the signal yields $N_{\rm sig}$ and the expected background yields $N_{\rm exp}$ in the signal region follow a Poisson distribution. We use the prior probability density function $1/\sqrt{N_{\rm sig} + N_{\rm exp}}$ to estimate the expected upper limits of signal yields based on a Bayesian approach~\cite{ZHU2007322}.
The uncertainty in $N_{\rm exp}$ and $\sigma$ are treated as two independent parameters when estimating the upper limits on the signal yields~\cite{ZHU2007322}.
The upper limits on the signal yields at 90\% credibility level are found to be 1.92 and 1.81, for the $\tau^- \rightarrow \Lambda \pi^-$ and $\tau^- \rightarrow \bar{\Lambda} \pi^-$ channels, respectively. The corresponding upper limits on the branching fractions are $4.7\times10^{-8}$ and $4.3\times10^{-8}$, respectively, while the expected upper limits based on background simulated samples are $7.2\times10^{-8}$ and $5.5\times10^{-8}$, respectively. If we use the flat prior probability density function, although it is not advised in Ref.~\cite{ZHU2007322}, the corresponding upper limits on the branching fractions will be $5.7\times10^{-8}$ and $5.5\times10^{-8}$.
These upper limits do not appreciably change if we do not include systematic uncertainties.


In summary, we present a search for the BNV and lepton number violation decays $\tau^- \rightarrow \Lambda \pi^-$ and $\tau^- \rightarrow \bar{\Lambda} \pi^-$ using a 364 fb$^{-1}$ data sample collected by the Belle II experiment at $\sqrt{s}$ = 10.58 GeV. No evidence of signal is observed, and upper limits at 90\% credibility level on the branching fractions $\BR(\tau^- \rightarrow \Lambda \pi^-)$ and $\BR(\tau^- \rightarrow \bar{\Lambda} \pi^-)$ are estimated to be $4.7 \times 10^{-8}$ and $4.3 \times 10^{-8}$, respectively. These are the most stringent constraints to date on the branching fraction of $\tau^- \rightarrow \Lambda \pi^-$ with $|\Delta(B-L)|=2$ and $\tau^- \rightarrow \bar{\Lambda} \pi^-$ with $|\Delta(B-L)|=0$. 


This work, based on data collected using the Belle II detector, which was built and commissioned prior to March 2019, was supported by
Higher Education and Science Committee of the Republic of Armenia Grant No.~23LCG-1C011;
Australian Research Council and Research Grants
No.~DP200101792, 
No.~DP210101900, 
No.~DP210102831, 
No.~DE220100462, 
No.~LE210100098, 
and
No.~LE230100085; 
Austrian Federal Ministry of Education, Science and Research,
Austrian Science Fund
No.~P~31361-N36
and
No.~J4625-N,
and
Horizon 2020 ERC Starting Grant No.~947006 ``InterLeptons'';
Natural Sciences and Engineering Research Council of Canada, Compute Canada and CANARIE;
National Key R\&D Program of China under Contract No.~2022YFA1601903,
National Natural Science Foundation of China and Research Grants
No.~11575017,
No.~11761141009,
No.~11705209,
No.~11975076,
No.~12135005,
No.~12150004,
No.~12161141008,
and
No.~12175041,
and Shandong Provincial Natural Science Foundation Project~ZR2022JQ02;
the Czech Science Foundation Grant No.~22-18469S;
European Research Council, Seventh Framework PIEF-GA-2013-622527,
Horizon 2020 ERC-Advanced Grants No.~267104 and No.~884719,
Horizon 2020 ERC-Consolidator Grant No.~819127,
Horizon 2020 Marie Sklodowska-Curie Grant Agreement No.~700525 ``NIOBE''
and
No.~101026516,
and
Horizon 2020 Marie Sklodowska-Curie RISE project JENNIFER2 Grant Agreement No.~822070 (European grants);
L'Institut National de Physique Nucl\'{e}aire et de Physique des Particules (IN2P3) du CNRS
and
L'Agence Nationale de la Recherche (ANR) under grant ANR-21-CE31-0009 (France);
BMBF, DFG, HGF, MPG, and AvH Foundation (Germany);
Department of Atomic Energy under Project Identification No.~RTI 4002,
Department of Science and Technology,
and
UPES SEED funding programs
No.~UPES/R\&D-SEED-INFRA/17052023/01 and
No.~UPES/R\&D-SOE/20062022/06 (India);
Israel Science Foundation Grant No.~2476/17,
U.S.-Israel Binational Science Foundation Grant No.~2016113, and
Israel Ministry of Science Grant No.~3-16543;
Istituto Nazionale di Fisica Nucleare and the Research Grants BELLE2;
Japan Society for the Promotion of Science, Grant-in-Aid for Scientific Research Grants
No.~16H03968,
No.~16H03993,
No.~16H06492,
No.~16K05323,
No.~17H01133,
No.~17H05405,
No.~18K03621,
No.~18H03710,
No.~18H05226,
No.~19H00682, 
No.~20H05850,
No.~20H05858,
No.~22H00144,
No.~22K14056,
No.~22K21347,
No.~23H05433,
No.~26220706,
and
No.~26400255,
the National Institute of Informatics, and Science Information NETwork 5 (SINET5), 
and
the Ministry of Education, Culture, Sports, Science, and Technology (MEXT) of Japan;  
National Research Foundation (NRF) of Korea Grants
No.~2016R1\-D1A1B\-02012900,
No.~2018R1\-A2B\-3003643,
No.~2018R1\-A6A1A\-06024970,
No.~2019R1\-I1A3A\-01058933,
No.~2021R1\-A6A1A\-03043957,
No.~2021R1\-F1A\-1060423,
No.~2021R1\-F1A\-1064008,
No.~2022R1\-A2C\-1003993,
and
No.~RS-2022-00197659,
Radiation Science Research Institute,
Foreign Large-Size Research Facility Application Supporting project,
the Global Science Experimental Data Hub Center of the Korea Institute of Science and Technology Information
and
KREONET/GLORIAD;
Universiti Malaya RU grant, Akademi Sains Malaysia, and Ministry of Education Malaysia;
Frontiers of Science Program Contracts
No.~FOINS-296,
No.~CB-221329,
No.~CB-236394,
No.~CB-254409,
and
No.~CB-180023, and SEP-CINVESTAV Research Grant No.~237 (Mexico);
the Polish Ministry of Science and Higher Education and the National Science Center;
the Ministry of Science and Higher Education of the Russian Federation
and
the HSE University Basic Research Program, Moscow;
University of Tabuk Research Grants
No.~S-0256-1438 and No.~S-0280-1439 (Saudi Arabia);
Slovenian Research Agency and Research Grants
No.~J1-9124
and
No.~P1-0135;
Agencia Estatal de Investigacion, Spain
Grant No.~RYC2020-029875-I
and
Generalitat Valenciana, Spain
Grant No.~CIDEGENT/2018/020;
National Science and Technology Council,
and
Ministry of Education (Taiwan);
Thailand Center of Excellence in Physics;
TUBITAK ULAKBIM (Turkey);
National Research Foundation of Ukraine, Project No.~2020.02/0257,
and
Ministry of Education and Science of Ukraine;
the U.S. National Science Foundation and Research Grants
No.~PHY-1913789 
and
No.~PHY-2111604, 
and the U.S. Department of Energy and Research Awards
No.~DE-AC06-76RLO1830, 
No.~DE-SC0007983, 
No.~DE-SC0009824, 
No.~DE-SC0009973, 
No.~DE-SC0010007, 
No.~DE-SC0010073, 
No.~DE-SC0010118, 
No.~DE-SC0010504, 
No.~DE-SC0011784, 
No.~DE-SC0012704, 
No.~DE-SC0019230, 
No.~DE-SC0021274, 
No.~DE-SC0021616, 
No.~DE-SC0022350, 
No.~DE-SC0023470; 
and
the Vietnam Academy of Science and Technology (VAST) under Grants
No.~NVCC.05.12/22-23
and
No.~DL0000.02/24-25.

These acknowledgements are not to be interpreted as an endorsement of any statement made
by any of our institutes, funding agencies, governments, or their representatives.

We thank the SuperKEKB team for delivering high-luminosity collisions;
the KEK cryogenics group for the efficient operation of the detector solenoid magnet;
the KEK computer group and the NII for on-site computing support and SINET6 network support;
and the raw-data centers at BNL, DESY, GridKa, IN2P3, INFN, and the University of Victoria for off-site computing support.


\begin{thebibliography}{**}
\bibitem{CL1} For Bayesian upper limits, it is now standard to quote
credibility intervals rather than confidence levels, which
are appropriate for frequentist analysis~\cite{CL2}.
\end{thebibliography}

\end{document}